\documentclass[twocolumn,pra,showpacs,preprintnumbers,amsmath,amssymb,superscriptaddress,floatfix]{revtex4}

\usepackage{latexsym}
\usepackage{epsfig}
\usepackage{graphicx}
\usepackage{textcomp}

\newcommand{\be}{\begin{equation}}
\newcommand{\ee}{\end{equation}}
\newcommand{\bea}{\begin{eqnarray}}
\newcommand{\beas}{\begin{eqnarray*}}
\newcommand{\eea}{\end{eqnarray}}
\newcommand{\eeas}{\end{eqnarray*}}
\newcommand{\ba}{\begin{array}}
\newcommand{\ea}{\end{array}}

\def\ls{\mathrel{\lower4pt\vbox{\lineskip=0pt\baselineskip=0pt
           \hbox{$<$}\hbox{$\sim$}}}}
\def\gs{\mathrel{\lower4pt\vbox{\lineskip=0pt\baselineskip=0pt
           \hbox{$>$}\hbox{$\sim$}}}}

\bibstyle{apsrev}
\begin{document}
\title{Non-degenerate four-wave mixing in rubidium vapor: transient regime}
\author{F. E. Becerra}
\affiliation{ Joint Quantum Institute, Department of Physics
University of Maryland and National Institute of Standards and
Technology, College Park, MD 20742 USA.} \affiliation{ Departamento
de F\'{\i}sica, CINVESTAV. Apdo. Post. 14-740, 07000, M\'exico,
D.F., M\'exico.}
\author{R. T. Willis}
\affiliation{ Joint Quantum Institute, Department of Physics
University of Maryland and National Institute of Standards and
Technology, College Park, MD 20742 USA.}
\author{S. L. Rolston}
\affiliation{ Joint Quantum Institute, Department of Physics
University of Maryland and National Institute of Standards and
Technology, College Park, MD 20742 USA.}
\author{H. J. Carmichael }
\affiliation{ Joint Quantum Institute, Department of Physics
University of Maryland and National Institute of Standards and
Technology, College Park, MD 20742 USA.}
\affiliation{Department of Physics, University of Auckland,
Private Bag 92019, Auckland, New Zealand.}
\author{L. A. Orozco }
\affiliation{ Joint Quantum Institute, Department of Physics
University of Maryland and National Institute of Standards and
Technology, College Park, MD 20742 USA.}


\begin{abstract}
We investigate the transient response of the generated light from Four-Wave Mixing (FWM) in the diamond configuration using a step-down field excitation.
The transients show fast decay times and oscillations that depend on the detunings and intensities of the fields. A simplified model taking into account
the thermal motion of the atoms, propagation, absorption and dispersion effects
shows qualitative agreement with the experimental observations with the energy levels in rubidium ($5S_{1/2}$, $5P_{1/2}$, $5P_{3/2}$ and $6S_{1/2}$).
The atomic polarization comes from all the contributions of different velocity classes of atoms in the ensemble modifying dramatically the total transient
behavior of the light from FWM.
\end{abstract}

\maketitle
\section{Introduction}\label{Introduction}

Four-Wave Mixing (FWM) is a non-linear optical process that coherently combines three waves to produce a fourth one, and is used in many areas of physics.
In atomic physics FWM is utilized to study spectroscopic and quantum optics processes
\cite{bloembergen64, demtroder81, schmitt97, boyd03, oudar80, slusher85}.
It can generate specific frequencies not available with direct electromagnetic sources, and has become a widely used tool for condensed
matter physics to characterize the short time correlations and coherences of semiconductors (see for example Refs. \cite{bloembergen64, shen02, shah99}).
Recent work with FWM for quantum information includes implementations of the atomic- ensemble-based quantum repeater of Duan, Lukin, Cirac, and Zoller \cite{duan01}
such as \cite{kuzmich03, van03, chaneliere06}, the generation of strong squeezing \cite{mccormick07, villar05} and quantum imaging \cite{boyer08}.
These implementations rely on the coherences created in the atomic systems, which are affected by
dephasing mechanisms producing decoherence over time. Study of the temporal dynamics of FWM can provide insight into the decay of the atomic polarization and the process
of decoherence.

\begin{figure}
\leavevmode \centering
  \includegraphics[width=2.8in]{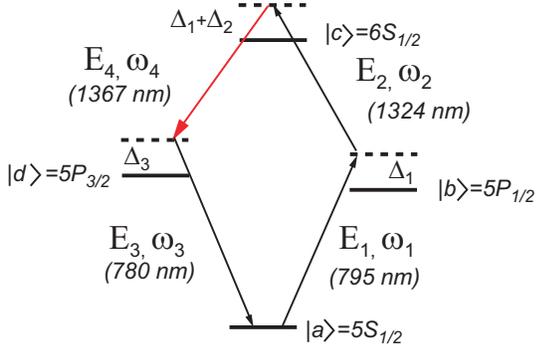}
  \renewcommand{\baselinestretch}{1}
\small\normalsize
\begin{quote}
  \caption{\label{AtomicConfigurationTran} (Color online) Four-level atom interacting with three external fields $E_{1}$, $E_{2}$ and $E_{3}$ with
frequencies $\omega_{1}$ (795 nm), $\omega_{2}$ (1324 nm) and $\omega_{3}$ (780 nm) respectively, and generated light (red online)  via four-wave mixing at frequency $\omega_{4}$ (1367 nm).
$E_{1}$ and $E_{3}$ are continuous wave (CW) fields and $E_{2}$ is time dependent.
The energy levels in the diamond configuration in $^{85}$Rb are $|a\rangle=5S_{1/2},~F=3$, $|b\rangle=5P_{1/2},~F=2$, $|c\rangle=6S_{1/2},~F=3$ and $|d\rangle=5P_{3/2},~F=4$.
  }
  \end{quote}
\end{figure}

We study the time evolution of the atomic coherences in FWM in rubidium vapor with transitions at telecommunication wavelengths.
We use a step-down excitation of the excited states and observe the generated light from FWM  in a diamond configuration, where
the levels include the 5S ground state, the 5P states (D1 and D2 lines), and the 6S state
( See Fig. \ref{AtomicConfigurationTran}).  This study complements
the investigation of the frequency response of the process in Refs. \cite{becerra08, willis09}
and concerns effects otherwise unobservable in the frequency regime due to the nonlinearity of the system.
The transient signal contains information about the temporal evolution
of the induced atomic polarization in the medium and can have collective atomic effects \cite{brownell95, lvovsky99}.
Since the atomic polarization is responsible for quantum effects \cite{duan01},
understanding the decoherence is fundamental for applications of FWM in quantum information;
our work \cite {willis07t2}
shows the limitations on non-classicality from decoherence.

The FWM signal in our configuration in steady state is weak. We step down to zero the laser intensity that connects
one of the intermediate states to the highest excited state and observe the time evolution of the FWM as it decays and
propagates out of the system. The time scales of this atomic system allow us good resolution with this simple step-down process;
this is in contrast to the situation encountered in FWM in semiconductors where the dynamics are very fast (picoseconds or faster)
and pump-probe techniques such as time-integrated FWM and time-resolved FWM are used to study correlation dynamics
in semiconductors \cite{axt04, chemla01, wang94, schafer96, wegener90}.

The organization of this work is as follows:
Sec. \ref{Theory} describes a simplified theoretical model of the system.
Sec. \ref{Experiment} contains the experimental setup and observations.
Sec. \ref{Analysis} contains the analysis incorporating other effects present in the experiment such as Doppler broadening, and
Sec. \ref{Results} explains the results of the investigation of the transients in FWM.
Sec. \ref{Conclusions} summarizes our conclusions.

\section{Theory}\label{Theory}
%
We study the evolution of the atomic coherences created by the non-linear interaction of three fields in the atomic medium
in the FWM process when one of those fields is suddenly set to zero.
The density matrix treatment allows us to study the decay and possible dephasing mechanisms such as Doppler broadening.

Figure  \ref{AtomicConfigurationTran} shows the  fields $E_{i}$ at frequencies $\omega_{i}$ coupling the four levels of the atomic system.
$E_{1}$ couples the transition $|a\rangle\rightarrow|b\rangle$, $E_{2}$ the transition $|b\rangle\rightarrow|c\rangle$ and $E_{3}$ the transition
$|a\rangle\rightarrow|d\rangle$. The generated light from FWM is resonant with the transition $|d\rangle\rightarrow|c\rangle$.
We assume that the generated light from the FWM process is weak in comparison with the incident fields and we ignore propagation effects at this frequency.

The density matrix equations describing this system in the rotating wave approximation (RWA) \cite{boyd03} are:
\begin{eqnarray}\label{DMequations}
\dot{\rho}_{ll}&=&-\sum_{n}\Gamma_{nl}\rho_{ll}+\sum_{n}\Gamma_{nl}\rho_{nn}\\ \nonumber
& & -\frac{i}{\hbar}\sum_{m}\left(V_{lm}\rho_{ml}-\rho_{lm}V_{ml}\right)\nonumber\\ \nonumber
\dot{\rho}_{mn}&=&(i\Delta_{mn}-\gamma_{mn})\rho_{mn}\\ \nonumber
 & & -\frac{i}{\hbar}\sum_{\nu}\left(V_{m\nu}\rho_{\nu n}-\rho_{m\nu}V_{\nu n}\right)
\end{eqnarray}
where $\rho_{ll}$ describes population of level $l$, $\rho_{mn}$ the coherence between the levels $m$ and $n$,  $\Gamma_{mn}$ the decay rate from level $m$ to $n$,
 $\Gamma_{n}=1/\tau_{n}$ the total decay rate of level $n$ with $\tau_{n}$ the lifetime of this level and  $\gamma_{mn}=\frac{1}{2}(\Gamma_{m}+\Gamma_{n})$ the dephasing
 rate of the coherence $\rho_{mn}$. The term $\Delta_{mn}$ is the detuning of the external field coupling the transition $|n\rangle\rightarrow|m\rangle$ from the energy
 of this transition (see Fig. \ref{AtomicConfigurationTran}). Energy conservation in FWM, $\omega_{1}+\omega_{2}=\omega_{3}+\omega_{4}$, defines the frequency and detuning
 for the generated light. The term $V_{nm}=-\hbar\Omega_{nm}$ describes the interaction of the atomic ensemble with the external field in the dipole approximation
 \cite{boyd03} where $\Omega_{nm}$ is the Rabi frequency for the external field coupling to the transition $|n\rangle\rightarrow|m\rangle$ given by
$\Omega_{i}=\vec{\mu_{i}}\cdot\vec{E}_{i}/\hbar$, where $\vec{\mu}_{i}$ is the electric dipole moment of the transition $|n\rangle\rightarrow|m\rangle$
and  $\vec{E}_{i}$ is the electric field coupling to this transition.

The density matrix equations are 16 coupled equations. Conservation of population, $\sum\rho_{nn}=1$, reduces this system to a set of 15 coupled differential equations.
The generated light from FWM resonant with the transition $|d\rangle\rightarrow|c\rangle$ (see Fig. \ref{AtomicConfigurationTran}) is proportional to the atomic
polarization oscillating at the frequency $\omega_{4}$ \cite{becerra08}. The induced atomic polarization for the transition $|d\rangle\rightarrow|c\rangle$ is
$P_{dc}=N\mu_{dc}\rho_{dc}$ where $N$ is the atomic number density \cite{allen72,boyd03}.
The study of the transient behavior of the light from FWM requires solving these coupled differential equations for the atomic coherence $\rho_{dc}$.

We assume that the system starts in the steady state at time $t=0$
when we turn off the field
$E_{2}$ (at $t=0$). The system evolves in the presence of only $E_{1}$ and $E_{3}$ with the steady state as the initial conditions. This allows some simplifications
in the calculation.
We obtain a decoupled subset of the system of 15 coupled equations that contains the term $\rho_{dc}$:
\begin{equation}
\frac{d}{dt}\left(
              \begin{array}{c}
                \rho_{ca} \\
                \rho_{cb} \\
                \rho_{dc} \\
              \end{array}
            \right)=\left(
                      \begin{array}{ccc}
                        \alpha_{ca} & i\Omega_{1} & i\Omega_{3} \\
                        i\Omega_{1} & \alpha_{cb} & 0 \\
                        i\Omega_{3} & 0 & \alpha_{dc} \\
                      \end{array}
                    \right)
                    \left(
              \begin{array}{c}
                \rho_{ca} \\
                \rho_{cb} \\
                \rho_{dc} \\
              \end{array}
            \right)
            \label{Tran3X3decopledE}
\end{equation}
where
$\alpha_{dc}=-i\Delta_{FWM}-\gamma_{cd}$, $\alpha_{ca}=-i(\Delta_{1}+\Delta_{2})-\gamma_{ca}$ and $\alpha_{cb}=-i\Delta_{2}-\gamma_{cb}$ with $\Delta_{i}$ the detuning of
the laser from the corresponding atomic transition, as shown in Fig. \ref{AtomicConfigurationTran}, and $\Delta_{FWM}=\Delta_{1}+\Delta_{2}-\Delta_{3}$. The steady state
solution of the density matrix equations determines the initial phases and amplitudes for these coherences. The solution for $\rho_{dc}$ is a linear combination of three
exponential functions of time with different characteristic energies (quasienergies),

\begin{equation}\label{TranY14}
\rho_{dc}=A_{1}e^{-i\lambda_{1}t}+A_{2}e^{-i\lambda_{2}t}+A_{3}e^{-i\lambda_{3}t}\\
\end{equation}
where $A_{j}$ $(j=1,2,3)$ are complex time-independent amplitudes and the quasienergies $\lambda_{j}$ are complex quantities whose real parts describe the oscillatory behavior
of the generated light and imaginary parts the exponential decay constants. These quantities depend on the detunings and the Rabi frequencies of the lasers, and the amplitudes
$A_{j}$  are determined from the initial (steady state) conditions of the system.

\begin{figure}
\leavevmode \centering
  \includegraphics[width=3in]{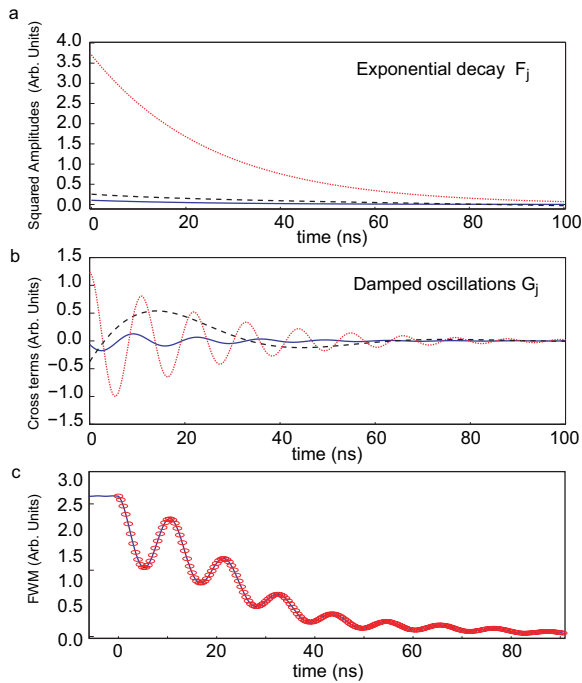}
  \renewcommand{\baselinestretch}{1}
\small\normalsize
\begin{quote}
  \caption{\label{Model_Osc_Damp_Compare}
  (Color online) FWM transients from the decoupled model.
  a) Square amplitudes F$_{j}$ from Eq. (\ref{DampTerms}). b) Damped oscillations G$_{j}$ from  Eq. (\ref{OscillTerms}).
  Dotted (red) lines correspond to the quasienergy $\lambda_{1}$, dashed (black) lines to $\lambda_{2}$ and solid (blue) lines to $\lambda_{3}$.
   c) Transient signal from the decoupled model using
    Eq. (\ref{TranY14}) (circles) and numerical simulation using the density matrix Eqs. (\ref{DMequations}) (solid line).
  Parameters used in the calculations:  $(\Delta_{1},\Delta_{2},\Delta_{3})=(0,270,30)$ MHz and Rabi frequencies of
$(\Omega_{1},\Omega_{2},\Omega_{3})=(50,10,10)$ MHz.
  }
  \end{quote}
\end{figure}

The real parts of the quasienergies show avoided crossings due to the interaction of the external fields with the atom.
Ref. \cite{becerra08} describes our investigations of the FWM resonant structure where we compare those quasienergies with the measured spectra. That work, using the Schrodinger
equation approach, does not take into account dissipation. With the density matrix formalism it is now possible to see the damping of the atomic polarization.
The intensity of the generated light from FWM is proportional to $|\rho_{dc}|^2$.
The quantity $|\rho_{dc}|^2$, using Eq. (\ref{TranY14}), contains three squared amplitudes associated with the same quasienergy,
\begin{equation}
F_{j}=|A_{j}|^2e^{-2{\rm{Im}}[\lambda_{j}]t},\qquad(j=1,2,3)
\label{DampTerms}
\end{equation}
showing exponential decay with a decay constant of $2\rm{Im}[\lambda_{j}]$, and three cross terms plus their complex conjugates (c. c.),
associated with different quasienergies,
\begin{equation}
G_{j}=A_{j}A_{k}^{*}e^{-i[\lambda_{j}-\lambda_{k}^{*}]t}+c.~c.,
\label{OscillTerms}
\end{equation}
for $(j,k=0,1,2),~j\neq k$. The latter show a damped oscillatory behavior with an oscillation frequency $\beta_{jk}=\rm{Re}[\lambda_{j}-\lambda_{k}^{*}]$ and an
exponential damping constant $\zeta_{jk}=\rm{Im}[\lambda_{j}-\lambda_{k}^{*}]$. Figs. \ref{Model_Osc_Damp_Compare}a and \ref{Model_Osc_Damp_Compare}b show the
quantities $F_{j}$ and $G_{j}$ as a function of time.
The calculations use the relevant parameters for the experiment. The relevant levels in $^{85}$Rb (Fig. \ref{AtomicConfigurationTran}) are
$|a\rangle=|5S_{1/2}\rangle$, $|b\rangle=|5P_{1/2}\rangle$, $|c\rangle=|6S_{1/2}\rangle$ and $|d\rangle=|5P_{3/2}\rangle$, with lifetimes $\tau_{5P_{1/2}}=27.8$ ns,
$\tau_{5P_{3/2}}=26.4$ ns and $\tau_{6S_{1/2}}=46$ ns and with corresponding linewidths
$\Gamma_{5P_{1/2}}=5.8$ MHz, $\Gamma_{5P_{3/2}}=6$ MHz and $\Gamma_{6S_{1/2}}=3.2$ MHz.
Figure \ref{Model_Osc_Damp_Compare}c shows the transient FWM light using the simplified model  from Eq. (\ref{TranY14}) and that numerically calculated from
the complete set of density matrix equations Eqs. (\ref{DMequations}).

For this particular set of Rabi frequencies and detunings, the transient signal of the generated field shows an oscillatory exponential decay behavior.
The oscillation frequency depends on the strength of the oscillatory terms from Eq. (\ref{OscillTerms}).
The decay of the total signal is dominated by the strongest components of the decaying terms in Eq. (\ref{DampTerms}).
The shape of the transient signal depends on these six terms with the phases of the oscillations determined by the initial conditions.
The oscillation frequencies depend on the detunings and the Rabi frequencies of the fields.

The decoupled subset of equations describes the behavior very well, as seen in Fig. \ref{Model_Osc_Damp_Compare}c, with the difference between
the two calculations of the order of one part in $10^{3}$, which is within the numerical error.
This allows us to study the physical mechanism acting in this system using the simpler set of equations.
The atomic coherences in the decoupled system can be represented as three coupled oscillators.
The oscillators corresponding to the atomic coherences $\rho_{dc}$ between $c$ and $d$ and $\rho_{cb}$ between $c$ and $b$
can store and exchange energy because of their mutual coupling. This exchange process depends on their relative phases.
The coherence $\rho_{ca}$ between $c$ and $a$ acts as part of the coupling coefficient in the system of two coupled oscillators $\rho_{dc}$ and $\rho_{cb}$.
Thus it produces a dynamic coupling between $\rho_{dc}$ and $\rho_{cb}$.

\section{Experimental observations of FWM transients}\label{Experiment}

The specific hyperfine levels that the fields couple in the experiment are such that
a field at 795 nm with frequency $\omega_{795}~(\omega_{1})$ couples the transition $5S_{1/2},~\mathord{F=3}\rightarrow5P_{1/2},~\mathord{F=2}$,  a field at 1324 nm
with frequency $\omega_{1324}~(\omega_{2})$ the transition $5P_{1/2},~\mathord{F=2}\rightarrow6S_{1/2},~\mathord{F=3}$ and a field at 780 nm with frequency $\omega_{780}~(\omega_{3})$
the transition $5S_{1/2},~\mathord{F=3}\rightarrow5P_{3/2},~\mathord{F=4}$ in  $^{85}$Rb (see Fig. \ref{AtomicConfigurationTran}). The generated light at
1367 nm from FWM is resonant with the transition $5P_{3/2},~\mathord{F=4}\rightarrow6S_{1/2},~\mathord{F=3}$ in $^{85}$Rb with a frequency $\omega_{1367}~(\omega_{4})$.
The estimated Rabi frequencies of the external excitations
(40$\%$ uncertainty due to their intensity profile) are $\Omega_{795}=45$ MHz, $\Omega_{1324}=8$ MHz and $\Omega_{780}=8$ MHz.

We use a fast fiber-coupled electro-optical modulator (EOM) to extinguish the excitation at 1324 nm in less than 4 ns, which is shorter than any atomic lifetime
of the levels involved in the FWM process.
A single photon detector collects the light from FWM as we turn off the 1324 nm light while the 780 nm and 795 nm light beams are operating in continuous wave (CW) mode.

\begin{figure}
\centering
\includegraphics[width=3in]{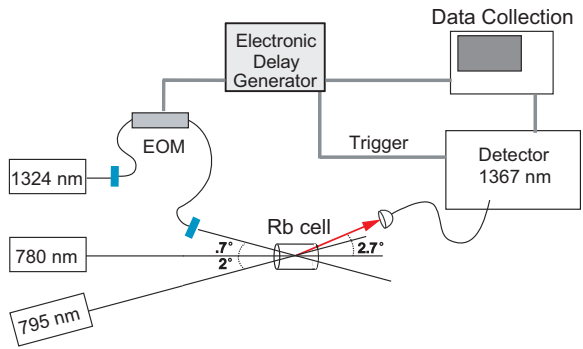}
\caption{\label{TransExpConf} (Color online) Schematic of the experimental configuration for the study of transient FWM. An electro-optic modulator (EOM)
produces a pulse of light at 1324 nm with a turn-off of less than 4 ns while two lasers at 780 nm and 795 nm work in continuous wave (CW) mode. A triggered
single photon detector detects FWM light as a function of delay from the excitation pulse. The angles among the beams satisfy phase matching for the FWM.
}
\end{figure}

Figure \ref{TransExpConf} shows the experimental configuration for the study of the transient regime of the FWM after the extinction of excitation light at 1324 nm.
This setup is similar to that used in the study of the process in the frequency domain described in Ref. \cite{becerra08}, but here we use a rubidium cell at 100$^{\circ}$C
containing enriched $^{85}$Rb (99$\%$ purity specified by the manufacturer) without buffer gas and without coating on the windows.
The cell is $L=$1.5 cm long and 2.5 cm in diameter.
The typical powers of the lasers are: P$_{795}$=2 mW, P$_{780}$=20 $\mu$W and the 1324 nm has a power of P$_{1324}$=0.8 mW in the CW mode after the fiber modulator.
All the beams have horizontal polarization as does the generated light from FWM.
The angles chosen to satisfy the phase-matching condition for the FWM process are shown in Fig. \ref{TransExpConf} \cite{becerra08}.
We stabilize the frequencies of the lasers at 780 nm and 795 nm using RF sidebands and saturated absorption to lock their frequencies using the absorption lines in rubidium.
We tune the frequencies of the lasers using acousto-optic modulators in the frequency locking setups. We use a transfer cavity lock \cite{zhao98} for the frequency
stabilization of the laser at 1324 nm.

A 10 Gbits/s Lithium Niobate Electro-Optic Modulator (EOM) from Alcatel-Lucent  with a bandwidth of 10 GHz modulates the amplitude of the 1324 nm light.
A digital delay generator  SRS DG535 produces a 300 ns pulse of 4 Volts and provides the AC signal driving the EOM. We use a DC power supply for the DC
voltage to the EOM for maximal suppression of light in the off stage of about $98\%\pm2\%$ at 1324 nm.

The Single Photon Benchtop Receiver from Princeton Lightwave, model PGA-600, is an In-GaAs based single photon avalanche photodiode (APD) running in a gated mode to detect the generated FWM light.
The gate has a 1 ns width which defines the temporal resolution in the experiment. We use optical attenuators to attenuate the light from FWM going into the fiber-coupled
detector to keep a maximal rate less than $1\times10^{-2}$ photons per 1 ns window. We use a FEL 1350 long-pass filter from Thorlabs before a single mode fiber (SMF) with transmission of
70$\%$ at 1367 nm and no appreciable transmission at 1324 nm. (We do not observe any counts caused by the 1324 nm laser when we trigger this detector at a rate of
$100\times10^3$ counts per second).
We use LabView programming and GPIB and serial communication to control the SRS DG535 and the PGA-600 for the generation of 1324 nm optical pulses and data
collection of the light from FWM.

\begin{figure}
\leavevmode \centering
  \includegraphics[width=2.9in]{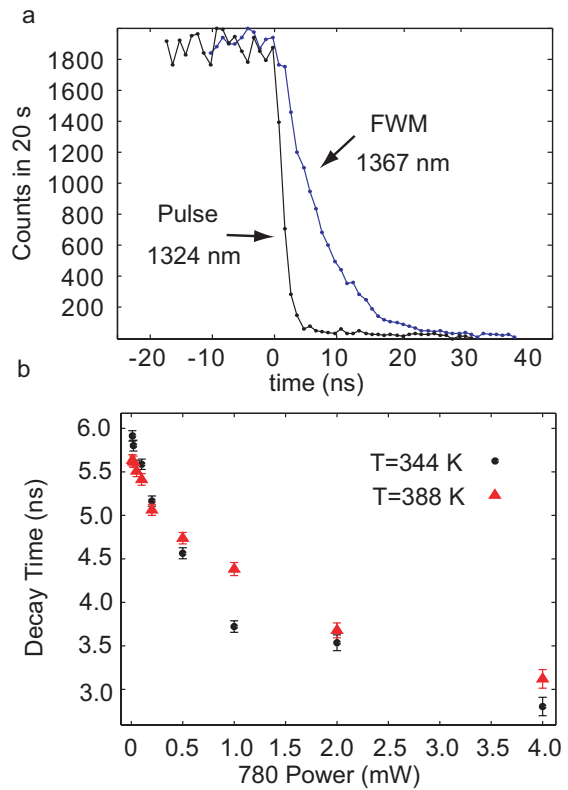}
  \renewcommand{\baselinestretch}{1}
\small\normalsize
\begin{quote}
  \caption{\label{Ttrans_Pulse_Decay} (Color online) a) Excitation pulse at 1324 nm (in black) and light from FWM at 1367 nm (in blue). b) Decay time of the transient
  FWM as a function of 780 nm power for temperatures T=344 K in (black) points and T=388 K in (red) triangles for P$_{795}=2$ mW. b has its vertical axis zero suppressed.
  }
  \end{quote}
\end{figure}

\begin{figure}
\leavevmode \centering
  \includegraphics[width=3.4in]{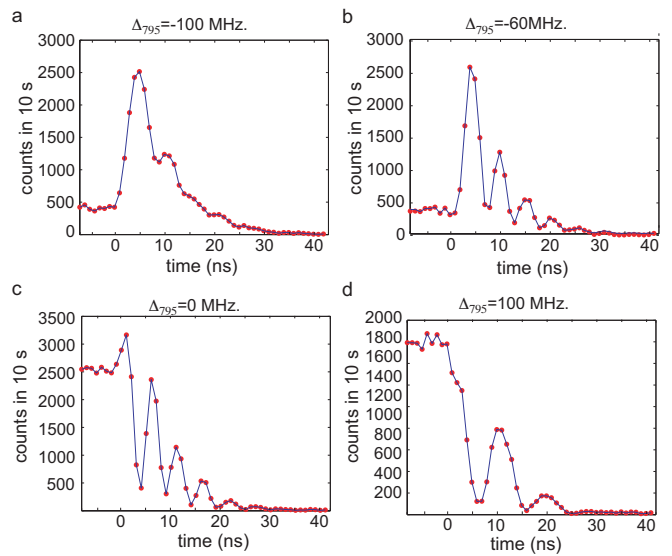}
  \renewcommand{\baselinestretch}{1}
\small\normalsize
\begin{quote}
  \caption{\label{Tran795Det} (Color online) Experimental FWM transients as a function  of 795 detuning for $\Delta_{780}=30$ MHz and $\Delta_{1324}$=270 MHz. a)
  $\Delta_{795}=-100$ MHz. b) $\Delta_{795}=-60$ MHz. c) $\Delta_{795}=0$ MHz. d) $\Delta_{795}=100$ MHz.
  The excitation pulse is extinguished at t=0 after the system has reached steady state. Continuous lines (blue) join the data points (red).
  }
  \end{quote}
\end{figure}

We study the transients in the light from FWM as a function of powers of the CW lasers and detunings of the excitation fields from specific atomic transitions
in $^{85}$Rb in a diamond configuration as shown in Fig. \ref{AtomicConfigurationTran}. 
The detunings and powers determine the exact values of the complex quasienergies that govern the atomic polarization (Eq. (\ref{TranY14})). 
We observe pure decay for certain regions of power and detunings (see Fig \ref{Ttrans_Pulse_Decay}) but also oscillatory decay for other regions 
(see Fig. \ref{Tran795Det}).

Figure \ref{Ttrans_Pulse_Decay}a shows the excitation light at 1324 nm (in black) and the light from FWM at 1367 nm (in blue)
as a function of time when the light at 1324 nm turns off at $t=0$ for the detunings $(\Delta_{795},\Delta_{1324},\Delta_{780})=(0,100,80)$ MHz
and laser powers $(P_{795},P_{1324},P_{780})=(2, 0.8, 0.02)$ mW.
Note the fast decay time of the light from FWM after the extinction of light at 1324 nm.
Fig. \ref{Ttrans_Pulse_Decay}b shows the decay time extracted from least squares fits to simple exponential
functions of the light from FWM as a function of the power of 780 nm light.
The errors are calculated from the change of the extracted parameter to increase the minimized $\chi^2$ of the fit by one.

We observe that the decay time decreases as we increase the power of the 780 nm light,
which might point to dephasing from power  broadening of the level $5P_{3/2}$ affecting the coherence $\rho_{dc}$.
This decay time is independent of atomic density in
a range of $7.89\pm2.51\times10^{11}$ to $1.5\pm.35\times10^{13}$ atoms/cm$^3$ corresponding to temperatures $T=344\pm5^\circ$K to $388\pm5^\circ$K
respectively as seen in Fig. \ref{Ttrans_Pulse_Decay}b. This suggests that the
short decay time is not due to any superradiant effects \cite{rehler71} or superfluorescence in the FWM process \cite{lvovsky99}.
The short decay time of the transient FWM can be caused by dephasing mechanisms related to atomic diffusion and Doppler broadening
in our sample \cite{brewer72}.
The decay does not depend on the power of the 795 nm light in a range from 20 $\mu$W to 2 mW.

When the external fields interact with the atomic system, they generate an atomic polarization grating in the atomic ensemble which can generate light from FWM \cite{shen86}.
This grating can diffuse due to the thermal motion of the atoms in the ensemble and contribute to dephasing in the process, shortening the decay time \cite{willis07t2}.
The study of this dephasing process as a function of thermal motion would require larger changes in the temperature of the ensemble than the temperature range used in
this experiment.

We study the transient behavior as a function of detunings of the three different lasers.
We use 20 $\mu$W of power for the 780 nm light and 2 mW of power for the 795 nm light.
The power of the 1324 laser in the CW mode, i.e., full transmission through the EOM, is 0.8 mW.
This set of powers results in a FWM transient signal with a relatively long decay time, $\tau_{Exp}\simeq6.0$ ns, as shown in
Fig. \ref{Ttrans_Pulse_Decay}b, and the maximum generation of FWM light. This allows for a more detailed study of the transients in FWM as a function of detunings.

Fig. \ref{Tran795Det} shows examples of the experimental observation of the light from FWM in the transient regime as a function of the detuning of the 795 light,
for $\Delta_{780}=30$ MHz and $\Delta_{1324}$=270 MHz.
%
We observe different features in the generated light: a FWM-spike as we turn off the 1324 nm for $\Delta_{795}=-100$ MHz and $\Delta_{795}=-60$ MHz, and oscillations
at different frequencies.
The FWM spike has an intensity above the steady state level, which corresponds to negative times in the figures. The decay times of the transients shown in
Fig. \ref{Tran795Det} are approximately 6 ns which is consistent with the decay times in Fig. \ref{Ttrans_Pulse_Decay}b for P$_{780}=20$ $\mu$W.
The oscillation frequency in FWM depends on the atomic polarization ( See Eq. (\ref{TranY14}) and Fig. \ref{Model_Osc_Damp_Compare})
This is different from quantum beats from the hyperfine structure \cite{willis07t2},
whose frequency remains mostly fixed to power and detuning changes.

This effect has been observed in the study of the exchange excitation of a system of N two-level atoms coupled to a cavity mode in the linear \cite{brecha95}
and nonlinear \cite{mielke97} regimes, where oscillations and spikes in the transient regime are the result of exchanging energy between the electromagnetic field
and induced atomic polarizations (vacuum Rabi oscillations \cite{raizen89, zhu90}). The asymmetry of the FWM-spike in our experiment comes from the phases in the
atomic polarizations that depend on the detunings and the intensities (Rabi frequencies) of the three fields at 795, 780 an 1324 nm.

We have explored the light from FWM  for different regimes of laser detunings,
and we have observed similar features in the transients
as those shown as a function of 795 detuning in Fig. \ref{Tran795Det}.

The parameter space in detuning and Rabi frequencies for FWM is large. Our exploration tries to highlight
three main features: damping faster than the lifetime of the transitions (Fig. \ref{Ttrans_Pulse_Decay}) when the detuning are large,
while for smaller detuning we look into the overshoot from energy stored in the atomic polarization (Fig. \ref{Tran795Det})
and the oscillations from the atomic polarization interference (Fig. \ref{Tran795Det}) or a combination of both.
At the end of section \ref{Results} we map out most of the region explored with respect to the predictions of the model of Eq. (\ref{TranY14}).



\section{Analysis}\label{Analysis}

When we consider a thermal atomic gas, the atoms will have velocities described by the Maxwell-Boltzmann distribution \cite{metcalf99}.
Different velocity classes of atoms will observe different frequencies of the external fields due to the Doppler effect \cite{metcalf99}.
The transient response in FWM depends on the detunings of the lasers and also on the velocity of the atoms in the ensemble.

The Doppler effect changes the transient behavior of the light dramatically when we consider the temperature of the  atomic ensemble.
We take into account the Doppler broadening by introducing velocity-dependent detunings, and
averaging the density matrix elements over the Maxwell-Boltzmann distribution \cite{metcalf99}
\begin{equation}
\rho_{mn}=\int\rho_{mn}(v)P(v)dv
\label{DoppAve}
\end{equation}
where $P(v)=\sqrt{m/2\pi k_{B}T}\exp{[-mv^2/(2k_{B}T)]}$ is the one-dimensional Maxwell-Boltzmann distribution with $m$ the atomic mass, $k_{B}$ the
Boltzmann constant and $T$ the temperature of the gas. The total generated light from FWM is
the result of contributions from all the different velocity classes in the ensemble.

\begin{figure}
\leavevmode \centering
  \includegraphics[width=3.3 in]{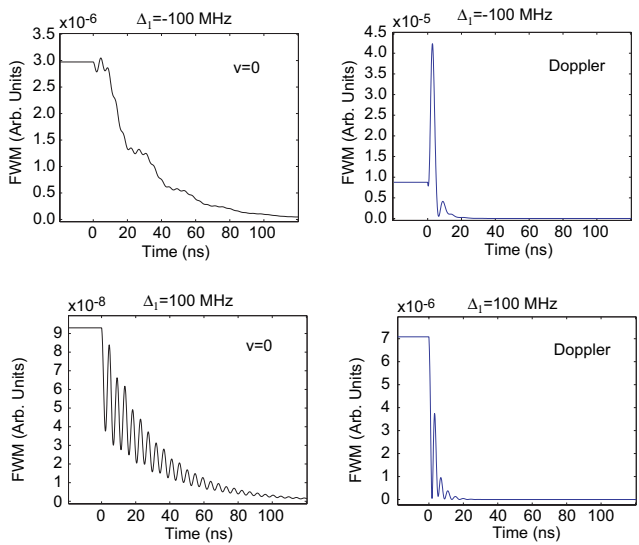}
  \renewcommand{\baselinestretch}{1}
\small\normalsize
\begin{quote}
  \caption{\label{TransDoppAve}
  (Color online) Calculation of the FWM light in the transient regime after $E_{2}$ is turned off at time t=0 as a function of $\Delta_{1}$ ($\Delta_{795}$),
  for $\Delta_{2}=270$ MHz, corresponding to $\Delta_{1324}$, and $\Delta_{3}=30$ MHz, corresponding to $\Delta_{780}$, with Rabi frequencies
  $(\Omega_{1},\Omega_{2},\Omega_{3})=(95,20,20)$ MHz. The left column corresponds to the Doppler-free case (\textit{v}=0); the right column after Doppler averaging.
  }
  \end{quote}
\end{figure}
Figure \ref{TransDoppAve} shows calculations of the transient behavior of the light from FWM for two detunings $\Delta_{1}$ ($\Delta_{795}$), $100$ MHz and
$-100$ MHz, with $\Delta_{2}=270$  MHz, and corresponding to $\Delta_{1324}$, $\Delta_{3}=30$ MHz, corresponding to $\Delta_{780}$, with Rabi frequencies
$(\Omega_{1},\Omega_{2},\Omega_{3})=(95,20,20)$ MHz.
The left column corresponds to zero atomic velocity; the right column corresponds to the case including the Doppler effect due to the motion of the atoms
at $T=$100$^{\circ}$C using Eq. (\ref{TranY14}) and the 1-dimensional (1-D) Doppler average from Eq. (\ref{DoppAve}).
The different velocity classes in the ensemble contribute to the light from FWM and modify its lineshape from the Doppler-free case.
Note the shortening of the decay time of the transients after the Doppler averaging.

The oscillation frequencies of the transient FWM should depend on the detunings of the lasers as well as on the strength of the Rabi frequencies
$\Omega_{1}$ and $\Omega_{3}$. They do not depend on the Rabi frequency $\Omega_{2}$ since it
does not enter into the decoupled equations, Eqs. (\ref{Tran3X3decopledE}), because it is off; $\Omega_{2}$ only contributes to the initial phases of the oscillations.
We use the steady state resonant structure of the FWM light from the experiment to estimate the Rabi frequency strengths.

We next consider absorption of the light at 780 and 795 nm from hyperfine levels in Rb, neglecting any absorption at 1324 nm. The absorption modifies the Rabi
frequencies entering into the process for different wavelengths of the external fields.


The linear absorption and dispersion of light in an atomic medium is described by the first order susceptibility $\chi^{(1)}$ where $P^{(1)}=\chi^{(1)}E$. For a two-level atom \cite{boyd03}:
\begin{equation}
\chi^{(1)}(\Delta)=A\frac{1}{\Delta-i\Gamma},
\label{chi1}
\end{equation}
where $\Delta$ is the detuning of the laser from the atomic transition, $\Gamma$ is the linewidth of the transition and $A$ is a constant.
We assume that the total absorption of light from different hyperfine levels can be described by a total linear susceptibility equal to the sum of the linear susceptibilities of each transition,
\begin{equation}
\chi^{(1)}_{Total}=\sum_{l}A_{l}\frac{1}{\Delta_{l}-i\Gamma},
\label{chi1Total}
\end{equation}
where $A_{l}$ defines the relative transition strengths of the transitions and $\Delta_{l}$ the detuning of the laser from each hyperfine state.
We account for the Doppler effect by averaging over velocities to obtain the total Doppler-averaged linear susceptibility $\bar{\chi}^{(1)}_{Total}$, where the bar denotes
Doppler-averaged. The field propagating in the medium is
\cite{allen72, boyd03}:
\begin{equation}
E(z)=E(0)\exp{(-{\frac{1}{2}\rm{OD}(z)\rm{~Im}}[\bar{\chi}^{(1)}_{Total}])}
\label{FieldPE}
\end{equation}
where $E(0)$ is the incident field at $z=0$, $E(z)$ is the field at a distance $z$ into the atomic medium, $\rm{Im}[\bar{\chi}^{(1)}_{Total}]$
is the imaginary part of the linear susceptibility and OD$(z)=2\pi N|\vec{\mu}_{i}|^2|\vec{k}_{i}|z/\hbar$ is the optical depth at $z$, where $\vec{k}_{i}$
is the wave vector of the field $E_{i}$ and $\vec\mu_{i}$ the dipole matrix element of the transition.




We find an OD=$20$
for light at 780 nm by comparing the observed absorption profile in the experiment with the theoretical Doppler-averaged
absorption profile from Eq. (\ref{FieldPE}) taking into account contributions from the other hyperfine levels. There is asymmetry in the absorption profile
around the detuning for the hyperfine level considered in the FWM process $5P_{3/2},F=4$. From the absorption data and the model we infer an experimental
OD for light at 795 nm of $9.8$.

\begin{figure}
\leavevmode \centering
  \includegraphics[width=3in]{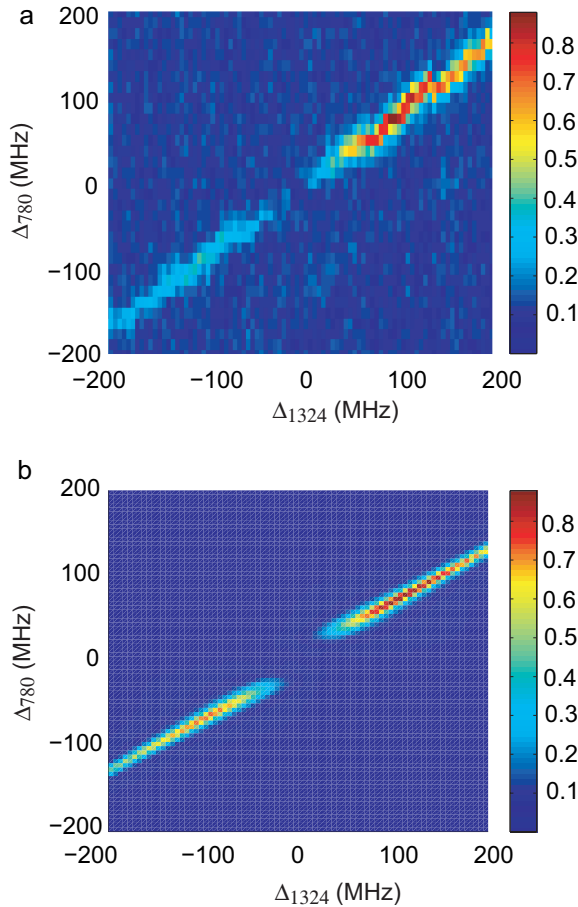}
  \renewcommand{\baselinestretch}{1}
\small\normalsize
\begin{quote}
  \caption{\label{TransFWM2D_Exp_Models}
  (Color online) Generated light from FWM in steady state as a function of detunings. a) Experimental FWM light as a function of 780 and 1324 detunings
  when the laser at 795 is locked on resonance. b) FWM light from
  the model taking into account
  Doppler broadening and propagation effects of the fields for incident Rabi frequencies $(\Omega_{1},\Omega_{2},\Omega_{3})=(45,7,7)$ MHz. Color bar
  in arbitrary units. Note the asymmetry in the resonant structure in both cases.
  }
  \end{quote}
\end{figure}
Figure \ref{TransFWM2D_Exp_Models}
shows the generated light from FWM in steady state as a function of the detunings $\Delta_{780}$ and $\Delta_{1324}$ when the 795 laser is on resonance.
Note the asymmetry in the intensity of the FWM light from the atomic resonances.

This asymmetry in the resonant structure shows that the Rabi frequencies entering into the FWM process are not symmetric with respect to zero detuning of the lasers.
This effect modifies the transient behavior of the light from FWM at different detunings.
The asymmetric behavior of the Rabi frequencies comes from the absorptive and dispersive properties of the medium due to hyperfine levels in the atomic structure of Rb.
The model (Fig. \ref{TransFWM2D_Exp_Models} b) taking into account Doppler effects and absorption shows a similar asymmetry in the resonant
structure of the light from FWM.


\section{Results}\label{Results}

The transient light from FWM in the experiment for different detunings shows a shortening of the decay time as theoretically predicted in Fig. \ref{TransDoppAve}.
This is due to induced atomic polarization interference in the atomic medium \cite{koch92},
in our particular case from different velocity classes of atoms in the Doppler-broadened atomic ensemble
as predicted by Zuo {\it et al.} \cite{zuo07}.
Different velocity classes experience different detunings and initial phases given by their steady state conditions. The total generated FWM field is the sum of all
the generated fields from different velocity classes. The FWM intensity, proportional to the square of the total field and thus to the square of the induced atomic polarization,
contains interference terms from different velocity classes.

\begin{figure}
\leavevmode \centering
  \includegraphics[width=2.7in]{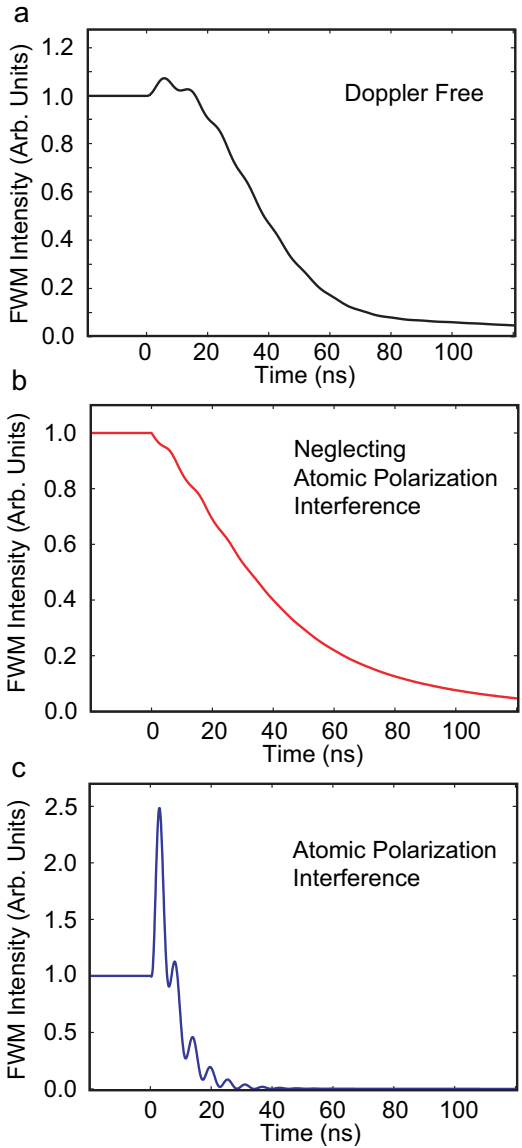}
  \renewcommand{\baselinestretch}{1}
\small\normalsize
\begin{quote}
  \caption{\label{PolarizationInter}
  (Color online) Theoretical calculation for Transient FWM for $(\Delta_{795},\Delta_{1324},\Delta_{780})=(-60,270,30)$ MHz and Rabi frequencies
  $(\Omega_{1},\Omega_{2},\Omega_{3})=(45,7,7)$.
  a) Zero-velocity class. b) Intensity contributions from all velocity classes excluding interference terms.
  c) Doppler average including interference terms.
  }
  \end{quote}
\end{figure}

Figure \ref{PolarizationInter} shows a calculation of the intensity of the light from FWM with detunings $(\Delta_{795},\Delta_{1324},\Delta_{780})=(-60,270,30)$ MHz
and Rabi frequencies $(\Omega_{1},\Omega_{2},\Omega_{3})=(45,7,7)$ MHz
for a) the zero-velocity class, b) the sum of Doppler-broadened intensities from different velocity classes neglecting the induced atomic polarization interference in the atomic medium using
$\int|\rho_{mn}(v)|^2P(v)dv$ and c) Doppler broadening including the induced atomic polarization interference calculated from the square of the Doppler-broadened atomic response using
Eq. (\ref{DoppAve}) $|\int\rho_{mn}(v)P(v)dv|^2$ \cite{zuo07}. The Rabi frequencies are consistent with those estimated from the FWM in the steady state regime (see Fig. \ref{TransFWM2D_Exp_Models}).

The transient signal ignoring induced atomic polarization interference in the atomic medium in Fig. \ref{PolarizationInter}b has a decay time, extracted from a fit of the
calculation to a simple exponential, of $22.31\pm3.8$ ns.
The transient signal containing induced atomic polarization interference in Fig. \ref{PolarizationInter}c shows a decay time of $5.82\pm1.09$ ns with a modulation frequency of $160\pm5.4$ MHz,
both numbers extracted from a fit to a simple exponential and from the peak frequency in a Fast Fourier Transform (FFT). This is consistent with the experimental decay time
of $\tau_{Exp}=6.3\pm0.09$ ns for the experimental Rabi frequencies shown in Fig. \ref{Tran795Det}b.
The shortening of the decay time of the transients with Doppler broadening corresponds to a broadening of the FWM spectrum.

The oscillations in the generated light are the result of the average of the oscillatory transients from different velocity classes of atoms.
The generation of light above the steady state level (FWM spike) is due to the total contribution to the FWM light from different velocity classes whose oscillations
phase after the 1324 nm light is turned off. At later times, after the in-phase interference,  the oscillations
get out of phase as the different induced atomic polarizations from different velocity classes evolve in time producing fast damping of the transients.
The model accounting for Doppler broadening qualitatively agrees with the experimental transients shown in Fig. \ref{Tran795Det} for specific detunings of the external
excitations.

\begin{figure}
\leavevmode \centering
  \includegraphics[width=3.0in]{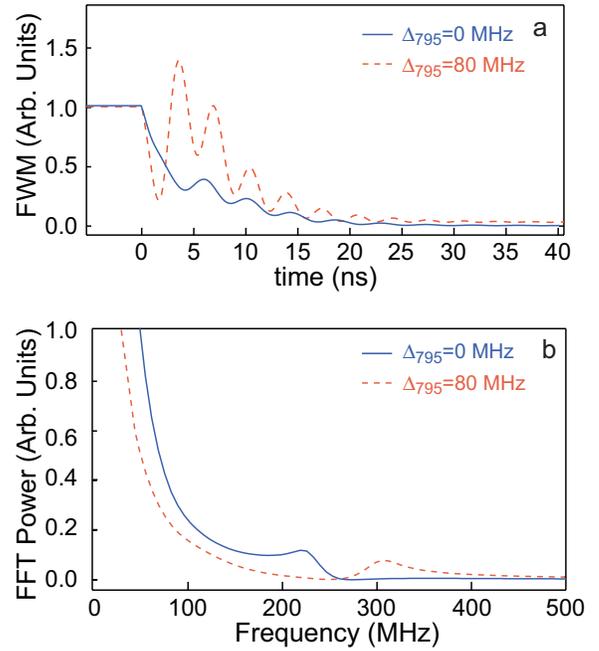}
  \renewcommand{\baselinestretch}{1}
\small\normalsize
\begin{quote}
  \caption{\label{TransFFTTheor} (Color online) Theoretical calculations of FWM transients Doppler-averaged  from  Eqs. (\ref{TranY14})  and (\ref{DoppAve})
  including pump depletion from
  Eq. (\ref{FieldPE}), as a function of detuning $\Delta_{795}$ for $(\Delta_{1324},\Delta_{780})=(270,30)$ MHz, together with their Fast Fourier Transforms (FFT).
  a) FWM transients for 795 nm resonant and non-resonant light in solid (blue) and dashed (red) lines, respectively.
  b) FFTs with frequency components at 250 MHz (HWHM=25 MHz) for resonant light and 295 MHz (HWHM=30 MHz) for
  non-resonant light in solid (blue) and dashed (red) lines, respectively.
  The Rabi frequencies $(\Omega_1; \Omega_2; \Omega_3) = (45; 7; 7)$ MHz are consistent with the experimental parameters.
  }
  \end{quote}
\end{figure}

Figure \ref{TransFFTTheor} shows examples of the modeled FWM transients taking into account Doppler broadening using Eq. (\ref{TranY14})
and pump depletion using Eq. (\ref{FieldPE}), together with the FFT showing their frequency components.

The oscillation frequency depends on the detunings of the lasers.
For detunings $(\Delta_{795},\Delta_{1324},\Delta_{780})=(0,270,30)$ MHz the transient FWM shows an oscillation frequency at
250 MHz with a half-width at half-maximum (HWHM) of 25 MHz.
For detunings $(\Delta_{795},\Delta_{1324},\Delta_{780})=(80,270,30)$ MHz the transient FWM shows an oscillation frequency at
295 MHz with HWHM=30 MHz.
The generated light from FWM contains contributions from different velocity classes of atoms resulting in an averaged oscillation frequency.
This dephasing mechanism  produces broadening as seen in Fig. \ref{TransFFTTheor} b and predicted in Ref. \cite{zuo07}.

\begin{figure}
\leavevmode \centering
  \includegraphics[width=3.1in]{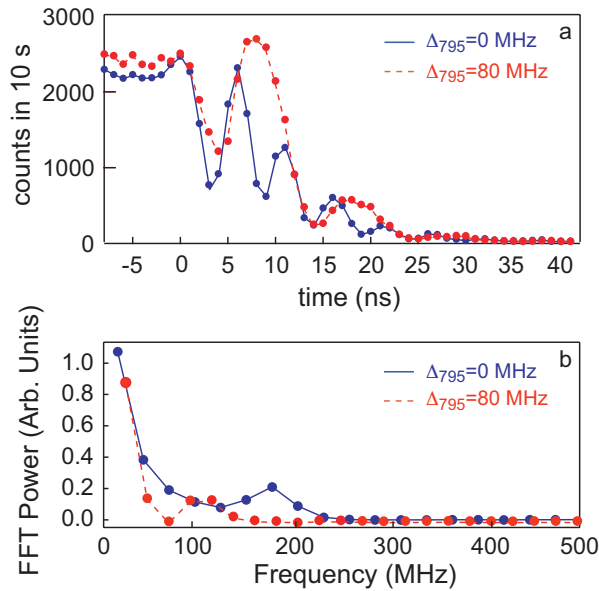}
  \renewcommand{\baselinestretch}{1}
\small\normalsize
\begin{quote}
  \caption{\label{TransFFTExp} (Color online) Experimental FWM transients as a function of detuning $\Delta_{795}$ for $(\Delta_{1324},\Delta_{780})=(270,30)$ MHz
  and FFTs for a situation similar to Fig.~\ref{TransFFTTheor}.
  a) FWM transients for 795 nm resonant and non-resonant light with data points joined by solid (blue) and dashed (red) lines, respectively.
  b) FFTs with frequency components at 185 MHz (HWHM=15 MHz) for resonant light and 100 MHz (HWHM=20 MHz) for
  non-resonant light with data points joined by solid (blue) and dashed (red) lines, respectively.
  }
  \end{quote}
\end{figure}

Figure \ref{TransFFTExp} shows measured examples of the light from FWM in the transient regime and the corresponding FFT. When the laser detunings are
$(\Delta_{795},\Delta_{1324},\Delta_{780})=(0,270,30)$ MHz the transient FWM contains oscillations with a frequency of 185 MHz (HWHM=15 MHz), and for
the detunings $(\Delta_{795},\Delta_{1324},\Delta_{780})=(80,270,30)$ MHz this oscillation changes to 100 MHz (HWHM=20 MHz)
as shown in Fig. \ref{TransFFTExp} b.

\begin{figure}
\leavevmode \centering
  \includegraphics[width=2.8in]{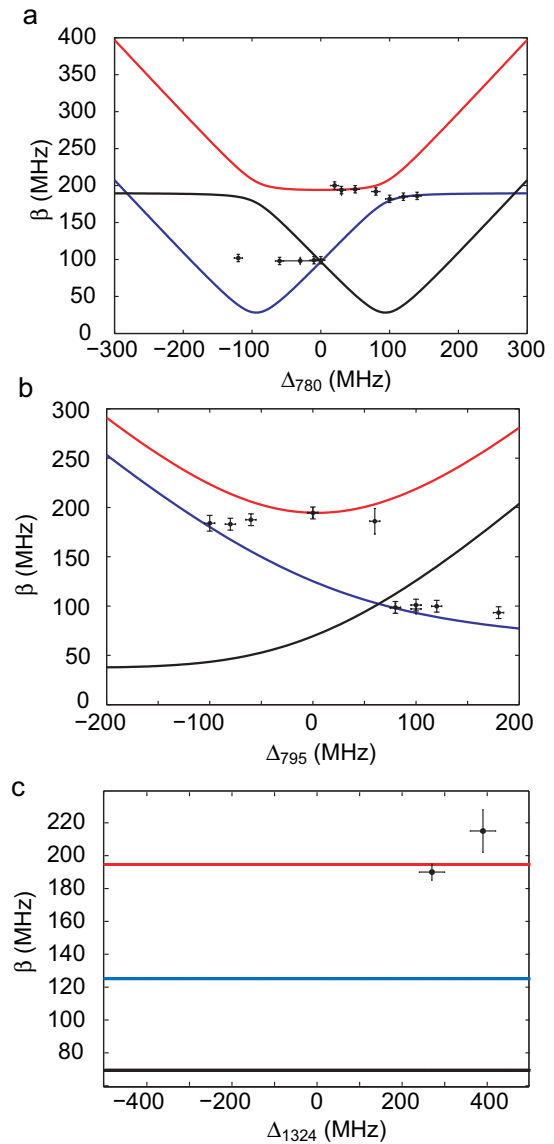}
  \renewcommand{\baselinestretch}{1}
\small\normalsize
\begin{quote}
  \caption{\label{BF_Data_Simulation} (Color online) Experimental (points with error bars) and modeled (continuous lines) oscillation frequencies
  $\beta_{jk}~(j\neq k)$ as a function of detunings. a) As a function of $\Delta_{780}$ with $(\Delta_{795},\Delta_{1324})=(0,270)$ MHz. b) As a function of
  $\Delta_{795}$ with $(\Delta_{780},\Delta_{1324})=(30,270)$ MHz. c) As a function of $\Delta_{1324}$ with $(\Delta_{780},\Delta_{795})=(30,0)$ MHz.
  The model ignores motion of the atoms.
  The Rabi frequencies $(\Omega_{1},\Omega_{2},\Omega_{3})=(95,20,20)$ MHz. The different colors of the lines correspond to oscillation frequencies from specific
  pairs of quasienergies in Eq. (\ref{OscillTerms}).
  }
  \end{quote}
\end{figure}

Figure \ref{BF_Data_Simulation} shows the oscillation frequencies $\beta_{j}$ from the decoupled model of Eq. (\ref{TranY14}) as a function of different
detunings and Rabi frequencies. The continuous lines are the results of the model and the points with error bars are measurements from FFTs under appropriate
experimental conditions. The horizontal error bars come from uncertainty in the frequencies of the lasers.
The lasers at 780 and 795 nm are locked to atomic absorption lines in Rb with linewidths of 6 MHz and frequency-shifted using optical modulators (see Sec. \ref{Experiment}).
We estimate the uncertainty of the frequency of the 1324 nm laser from the 30 MHz resolution of the wavemeter.

This simplified model shows some characteristics of the generated light. The transients should oscillate at the natural frequencies of the
induced atomic polarization in the medium, however, we observe some discrepancy between the experimental observations and the expected frequencies from the model.
These discrepancies could come from the fact that this simplified model does not contain information about pump absorption or transverse effects. Pump depletion could produce
abrupt changes in the Rabi frequencies entering into the FWM process as the lasers are scanned across atomic resonances because of the presence of other hyperfine levels.
Transverse effects in the atomic medium could modify the wave vectors of the pumps as a function
of frequency and change the phase matching condition required in FWM \cite{boyd03}.

The asymmetry with respect to the 795 and 780 detunings in the experimental oscillation frequencies
might be related to the asymmetric behavior of the observed FWM light in the steady state in Fig. \ref{TransFWM2D_Exp_Models}a.
The asymmetric absorption of the pump beams around the particular hyperfine levels predicts an asymmetry in the FWM in the steady state regime
(see Fig. \ref{TransFWM2D_Exp_Models}b).
However, a model fully describing the  asymmetric behavior of the transient oscillations in Fig. \ref{BF_Data_Simulation} might require solving the coupled
density matrix and field propagation equations to account for possible coupling between the pump fields in the medium producing focusing and defocusing as
well as nonlinear refractive index effects modifying the phase matching for the fields \cite{boyd03}.

The location of the oscillations (data points) shows that the system changes among the different oscillation frequencies available. The data comes from the
FFT of the time response and one oscillation dominates in every case. Further investigation will be necessary to understand this effect in the FWM process.

\section{Conclusions}\label{Conclusions}
%


Coherences, such as those created in the FWM process, are delicate. This paper shows the atomic polarization coherences that contribute to the FWM
and are visible through the time response of the system. They are responsible for transients, but their fragility is evident in the qualitative
and quantitative behaviors that strongly depend on detunings and intensities.
Our work also shows that the presence of absorption and dispersion with the Doppler-broadened medium significantly changes the coherences and their dynamics:
oscillations and decay time.


The observations of pure decay agree with the model for resonant excitation and low power; changes in the detuning bring oscillations and jumps above the
steady state value of the FWM signal which come from induced atomic polarization interference in the atomic medium.

The total transients result from contributions of different velocity classes of atoms in the Doppler-broadened medium; this produces induced atomic polarization
interference generated from different velocities broadening the FWM spectrum.
The model predictions for the amplitudes of the transients in FWM are in qualitative agrement with the experimental observations. The frequency structure,
related to the quasienergies of the system, is closer quantitatively to the model.

This work provides a better understanding of coherent effects and energy exchange processes in multilevel atomic systems and how they are affected
due to broadening as a source of decoherence. We find that, in this Doppler-broadened system, it is possible to obtain coherent processes with decay times
of the order of the radiative lifetimes for transitions in the diamond configuration.

\section*{Acknowledgments}
This work was supported by NSF, CONACYT and the Marsden fund of RSNZ.




\end{document}